\documentclass[superscriptaddress, nofootinbib, longbibliography, aps,pra, 11pt]{revtex4-2}
\usepackage[margin=1in,letterpaper]{geometry} 
\usepackage{tabularx} 
\usepackage{amsmath}  
\usepackage{graphicx} 
\usepackage{wrapfig}
\usepackage{amsthm}
\usepackage{amssymb}
\usepackage{bbm}
\usepackage{physics}
\usepackage[normalem]{ulem}
\usepackage{adjustbox}
\usepackage{comment}
\usepackage[section]{placeins}
\usepackage{mathrsfs}
\usepackage{mathtools}
\usepackage{natbib}
\usepackage{hyperref} 
\usepackage{lineno}
\usepackage{setspace}
\usepackage{times}
\usepackage{color}

\newtheorem{theorem}{Theorem}

\hypersetup{
	colorlinks=true,       
	linkcolor=blue,        
	citecolor=blue,        
	filecolor=magenta,     
	urlcolor=blue         
}

\newcommand{\eq}[1]{\text{Eq.~(\ref{eq:#1})}}
\newcommand{\fig}[1]{\text{Fig.~\ref{fig:#1}}}

\newcommand{\sect}[1]{Section~\ref{sec:#1}}

\newcommand{\create}[2]{#1^{\dag}_{#2}}
\newcommand{\ann}[2]{#1_{#2}}
\newcommand{\tsub}[2]{#1_{\text{#2}}}
\newcommand{\xgen}[3]{#1^{\dag}_{#2} #1_{#3} + #1^{\dag}_{#3} #1_{#2}}
\newcommand{\ygen}[3]{#1^{\dag}_{#2} #1_{#3} - #1^{\dag}_{#3} #1_{#2}}

\begin{document}
\title{Learning interacting fermionic Hamiltonians at the Heisenberg limit}

\author{Arjun Mirani}
\email{asmirani@stanford.edu}
\affiliation{Stanford Institute for Theoretical Physics, Stanford University, Stanford, CA 94305}
\affiliation{Department of Applied Physics, Stanford University, Stanford, CA 94305}
\author{Patrick Hayden}
\email{phayden@stanford.edu}
\affiliation{Stanford Institute for Theoretical Physics, Stanford University, Stanford, CA 94305}
\affiliation{Department of Physics, Stanford University, Stanford, CA 94305}

%\linenumbers

\begin{abstract}
Efficiently learning an unknown Hamiltonian given access to its dynamics is a problem of interest for quantum metrology, many-body physics and machine learning. A fundamental question is whether learning can be performed at the Heisenberg limit, where the Hamiltonian evolution time scales inversely with the error, $\varepsilon$, in the reconstructed parameters. The Heisenberg limit has previously been shown to be achievable for certain classes of qubit and bosonic Hamiltonians. Most recently, a Heisenberg-limited learning algorithm was proposed for a simplified class of fermionic Hubbard Hamiltonians restricted to real hopping amplitudes and zero chemical potential at all sites, along with on-site interactions. In this work, we provide an algorithm to learn a more general class of fermionic Hubbard Hamiltonians at the Heisenberg limit, allowing complex hopping amplitudes and nonzero chemical potentials in addition to the on-site interactions, thereby including several models of physical interest. The required evolution time across all experiments in our protocol is $\order{1/\varepsilon}$ and the number of experiments required to learn all the Hamiltonian parameters is $\order{\text{polylog}(1/\varepsilon)}$, independent of system size as long as each fermionic mode interacts with $\order{1}$ other modes. Unlike prior algorithms for bosonic and fermionic Hamiltonians, to obey fermionic parity superselection constraints in our more general setting, our protocol utilizes $\order{N}$ ancillary fermionic modes, where $N$ is the system size. Each experiment involves preparing fermionic Gaussian states, interleaving time evolution with fermionic linear optics unitaries, and performing local occupation number measurements on the fermionic modes. The protocol is robust to a constant amount of state preparation and measurement error.
\end{abstract}

\maketitle

\section{Introduction}
The problem of learning an unknown Hamiltonian, given access to its dynamics with time-evolution as a black-box operator, is of interest to a range of fields including quantum metrology \citep{Dutt_2023, Baumgratz_2016, Pang_2014, Ferrie_2012, Sergeevich_2011, Valencia_2004, Leibfried_2004, de_Burgh_2005, Lee_2002, McKenzie_2002, Bollinger_1996, Holland_1993, Wineland_1992, Caves_1981}, many-body physics \citep{Holzapfel_2015, Wang_2017, Qi_2019, Wiebe_2014_a, Wiebe_2014_b, Shabani_2011, Zhang_2014, Evans_2019, Li_2020, Zubida_2021, Rattacaso_2023, Yu_2023, Burgarth_2011, Franca_2022, Anshu_2021, Haah_2023, Sbahi_2022, Verdon_2019, Kwon_2020, Wang_2019, Huang_2020, Huang_qubit_2023, Li_boson_2023, Dutkiewicz_2023} and machine learning \citep{Anshu_2021, Haah_2023, Sbahi_2022, Verdon_2019, Kwon_2020, Wang_2019, Huang_2020}. The Hamiltonian learning problem can be considered a special case of quantum process tomography \citep{NielsenChuang}, asking whether a black-box physical process can be characterized up to some desired precision. Hamiltonian learning is increasingly of practical interest due to synthetic quantum systems and devices that need to be benchmarked \citep{Valenti_2019, Wiebe_2014_b, Wiebe_2015, hangleiter2024robustly, Da_Silva_2011}.

Investigating the \textit{efficiency} of Hamiltonian learning protocols is of both fundamental and practical interest. For instance, how much total time evolution $\textit{t}$, or how many copies $n$ of an entangled probe quantum state, are required to learn the parameters of the Hamiltonian to error $\varepsilon$? The fundamental quantum limit, representing the most efficient scaling, is the so-called `Heisenberg limit', which is either $t \sim \mathcal{O}(1/\varepsilon)$ or $n \sim \mathcal{O}(1/\varepsilon)$,  depending on whether time or entanglement is the metrological resource \citep{RPE_Kimmel_2015, RPE_Russo_2021, RPE_Belliardo_2020}. In the context of many-body Hamiltonians, most prior work (such as \citep{Wiebe_2014_b, Shabani_2011, Zhang_2014, Evans_2019, Li_2020, Zubida_2021, Rattacaso_2023, Yu_2023, Burgarth_2011, Franca_2022, Anshu_2021, Haah_2023}) achieves the so-called `standard quantum limit', where the total evolution time scales as $\Omega(1/\varepsilon^2)$. Only in the past two years has it been shown that the Heisenberg limit is achievable for certain classes of many-body Hamiltonians governing systems of qubits (first by \citep{Huang_qubit_2023}, followed by \citep{Dutkiewicz_2023}) and bosons \citep{Li_boson_2023}. The qubit algorithm of \citep{Huang_qubit_2023}, the first to achieve the Heisenberg limit in the many-body setting, introduced a Hamiltonian reshaping technique based on \textit{discrete quantum control}, which refers to interleaving time evolution with unitary gates. Their algorithm used this technique to effectively decouple the many-body Hamiltonian into multiple non-interacting clusters, allowing it to be learned in parallel via a divide-and-conquer approach, a technique subsequently adapted by \citep{Li_boson_2023} for the bosonic case. An alternative to discrete control is \textit{continuous} quantum control, which refers to continuously time-evolving the system under a modified Hamiltonian, where known terms are added to the original unknown Hamiltonian. The Heisenberg-limited qubit algorithm of \citep{Dutkiewicz_2023} uses continuous quantum control, allowing estimates of the Hamiltonian parameters to be adaptively refined through an iterative process. Notably, \citep{Dutkiewicz_2023} showed that for a large set of many-body Hamiltonians, including those that thermalize via the eigenstate thermalization hypothesis, achieving the Heisenberg limit requires quantum control, whether discrete or continuous.

Recently, \citep{Ni_fermion} proposed a Heisenberg-limited learning algorithm for a simplified subset of \textit{fermionic} Hubbard Hamiltonians restricted to real hopping amplitudes and zero chemical potential at all sites, along with on-site interactions.\footnote{Our research was performed independently. We only became aware of \citep{Ni_fermion} in the final stages of preparing this manuscript.} In this work, we provide an algorithm to learn a more general class of fermionic Hubbard Hamiltonians at the Heisenberg limit, allowing complex hopping amplitudes and nonzero chemical potentials in addition to the on-site interactions, thereby including several systems of physical interest. The approach  of \citep{Ni_fermion} is similar to ours, since both works adapt the technique of discrete quantum control used in the qubit \citep{Huang_qubit_2023} and bosonic \citep{Li_boson_2023} settings. However, generalizing to the complex hopping amplitudes and nonzero chemical potentials of our model requires additional ingredients to obey fermionic parity superselection constraints and learn the complex coefficients. These generalizations are physically relevant $-$ for instance, complex amplitudes are used to incorporate the effects of magnetic fields \citep{Gorg_2020, Ceven_2022}, and nontrivial chemical potential plays a key role in the properties of metals \citep{Mahan_book}. In the disordered setting, the variation of local energy across different sites, captured by the chemical potential coefficients, contributes to phenomena such as Anderson localization; the special case of of our model in which hopping terms and local energies are assigned randomly is known as the Anderson-Hubbard model~\citep{ulmke1995anderson,giovanni2021anderson}. Our protocol thus extends the regime of achievability of Heisenberg-limited learning to a physically well-motivated class of fermionic Hamiltonians. Such learning algorithms could potentially play a practical role in characterizing fermionic systems in the lab and benchmarking fermionic analog quantum simulators, which offer exciting possibilities to investigate condensed matter phenomena in regimes where classical computation is challenging \citep{Cheuk_2015, Haller_2015, Brown_2017, Hartke_2020, Koepsell_2020, Kale_2022, Xu_2023, lebrat2023observation}.

Our protocol consists of a series of experiments, each of which involves parallel preparation of two-mode fermionic Gaussian states that couple modes of the system to fermionic ancilla modes. This is followed by interleaving time evolution with fermionic linear optics (FLO) unitaries, and performing local occupation number measurements on the fermionic modes. The measurement results are efficiently post-processed classically according to the Heisenberg-limited Robust Phase Estimation (RPE) algorithm \citep{RPE_Kimmel_2015, RPE_Russo_2021, RPE_Belliardo_2020}. Overall, the protocol utilizes $\mathcal{O}(N)$ ancillae, where $N$ is the system size, although many of the experiments require no ancillae at all. The ancillae are used to prepare fixed-parity initial states of the form desired for RPE, as required by the physical constraint of fermionic parity superselection, explained in \sect{single_site_oscillator}. The total time evolution across all experiments scales as $\mathcal{O}(1/\varepsilon)$ and the number of experiments scales as $\mathcal{O}(\text{polylog}(1/\varepsilon))$. As long as the degree of the graph of interactions is bounded, these complexities are independent of the system size. Furthermore, our protocol is robust to a constant amount of state preparation and measurement error, a feature inherited from the robustness of RPE.

The rest of this paper is structured as follows: \sect{model_and_results} specifies the class of fermionic Hamiltonians under consideration and provides a statement of results. Sections \ref{sec:single_site_oscillator} to \ref{sec:many_body_H} describe the learning algorithm in a step-by-step manner, each section describing a subroutine that is used by the next, culminating in a divide-and-conquer approach to learning the full many-body Hamiltonian. \sect{discussion} concludes with a summary and discussion of future directions. Finally, Appendices \ref{appendix:app_a} and \ref{appendix:app_b} discuss the boundedness of the errors in the reconstructed parameters of the Hamiltonian.

\section{Results}\label{sec:results}

\subsection{Model and statement of results}\label{sec:model_and_results}
\begin{figure}[h]
\centering
\includegraphics[width = 0.7\linewidth]{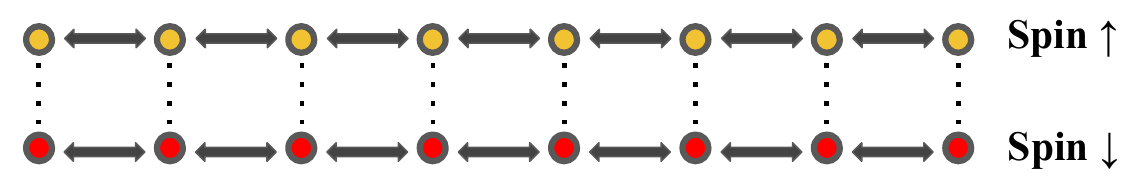}
\caption{Diagrammatic representation of a fermionic Hubbard Hamiltonian defined on a one-dimensional lattice. The yellow and red modes correspond to opposite spins. Dotted lines represent on-site repulsion between modes of opposite spin with the same spatial index. Arrows represent hopping interactions between neighboring spatial sites. For Hamiltonians defined on a more general graph, any two sites linked by an edge have a hopping interaction.}
\label{fig:Hamiltonian_def}
\end{figure}

The interacting fermionic Hamiltonians considered in this paper are of the following form, representing the Hubbard model:
\begin{equation}\label{eq:generic_H}
    H = \sum_{<i,j>} \sum_{\sigma \in \{\uparrow, \downarrow\}} h_{ij \sigma} \create{a}{i \sigma}\ann{a}{j \sigma} +  \sum_{\sigma \in \{\uparrow, \downarrow\}} \omega_{i\sigma} n_{i\sigma} + \sum_i \xi_i n_{i \uparrow} n_{i \downarrow}
\end{equation}
The indices $i,j$ denote spatial sites, corresponding to the vertices of a bounded-degree graph. Each spatial site comprises two spin modes with opposite spins ($\uparrow$ and $\downarrow$). The pair $\langle i,j \rangle$ denotes vertices connected by an edge in the underlying graph, generalizing the nearest-neighbor relation on a lattice. The operators $\create{a}{i \sigma}$ and $\ann{a}{i \sigma}$ are respectively the creation and annihilation operators for the fermionic spin mode at site $i$ with spin $\sigma$. They satisfy the canonical anticommutation relations: $\{\ann{a}{i \sigma}, \create{a}{j \rho} \}  = \delta_{ij}\delta_{\sigma \rho}$, while all other anticommutators vanish. The operator $n_{i \sigma} = \create{a}{i \sigma}\ann{a}{i \sigma}$ is the number operator for the corresponding spin mode. Due to the hermiticity of the Hamiltonian, the coefficients $h^*_{ij \sigma} = h_{ji \sigma}$ while $\omega_{i\sigma}$ and $\xi_i$ are real. We use similar notation to \citep{Li_boson_2023} for the coefficients to highlight the analogy to the bosonic case. In physical terms, the $h_{ij \sigma}$ coefficients are hopping amplitudes, the $\omega_{i\sigma}$ coefficients are chemical potentials, and the $\xi_i$ coefficients represent on-site interaction strength~\citep{fazekas1999lecture}. A one-dimensional example of a Hamiltonian of the form in \eq{generic_H} is depicted in \fig{Hamiltonian_def}.

Given access to black box time evolution generated by a Hamiltonian of the form in \eq{generic_H}, our goal is to learn the parameters $h_{ij\sigma}$, $\omega_{i \sigma}$ and $\xi_i$ for all the fermionic modes, with root-mean-square (RMS) error upper bounded by some $\varepsilon$. In Sections \ref{sec:single_site_oscillator} to \ref{sec:many_body_H} below, we describe a protocol that achieves this with Heisenberg scaling, thereby establishing the following result:
\begin{theorem}\label{main_thm}
Given the ability to
\begin{enumerate}
    \item prepare two-mode fermionic Gaussian states of the system and ancilla modes,
    \item apply black box time evolution under the unknown Hamiltonian,
    \item apply specified two-mode unitaries of the fermionic linear optics (FLO) form, and 
    \item perform local occupation measurements on the system and ancilla modes,
\end{enumerate}
the set of coefficients $\mathcal{C} \coloneq \{ h_{ij\sigma}, \omega_{i \sigma}, \xi_i \, | \, \forall i,j \! \in \! [N],\sigma \! \in \! \{\uparrow, \downarrow\}\}$ of a Hamiltonian of the form in \eq{generic_H} can be learned such that, for the estimator $\hat{\lambda}$ of any coefficient $\lambda \in \mathcal{C}$, the root-mean-square error $\sqrt{\mathbb{E}[(\hat{\lambda}-\lambda)^2]} \text{ is bounded above by } \varepsilon$. The protocol requires resources scaling as follows:\footnote{Here big-$\mathcal{O}$ notation means that there is a constant prefactor independent of $N$ and $\varepsilon$.}
\begin{enumerate}
    \item total amount of evolution time = $\mathcal{O}(1/\varepsilon)$
    \item total number of experiments = $\mathcal{O}(\text{polylog}(1/\varepsilon))$
    \item total number of ancillae = $\mathcal{O}(N)$
    \item total number of FLO unitaries = $\mathcal{O}(N \varepsilon^{-2}\text{polylog}(1/\varepsilon))$
\end{enumerate}
Additionally, the protocol is robust to a constant amount of state preparation and measurement (SPAM) error.
\end{theorem}

The rest of the paper constitutes a demonstration of this theorem. We will begin by discussing the simplest example of our class of Hamiltonians in \sect{single_site_oscillator} $-$ namely, a fermionic quantum anharmonic oscillator Hamiltonian with a single spatial site. Next, we will consider a two-site coupled anharmonic oscillator Hamiltonian in \sect{two_coupled_oscillators}, and show how this can be reduced to the single-site case. Using these results, in \sect{many_body_H} we show how the many-body Hamiltonians described by \eq{generic_H} can be reduced to the two-site case, facilitating parallelized learning of the Hamiltonian coefficients. Appendices \ref{appendix:app_a} and \ref{appendix:app_b} discuss the boundedness of the errors in the reconstructed parameters of the Hamiltonian.

\subsection{Learning a single-site (two-mode) Hamiltonian}\label{sec:single_site_oscillator}
\begin{wrapfigure}{r}{0.25\textwidth}
    \centering
    \includegraphics[width=0.20\textwidth]{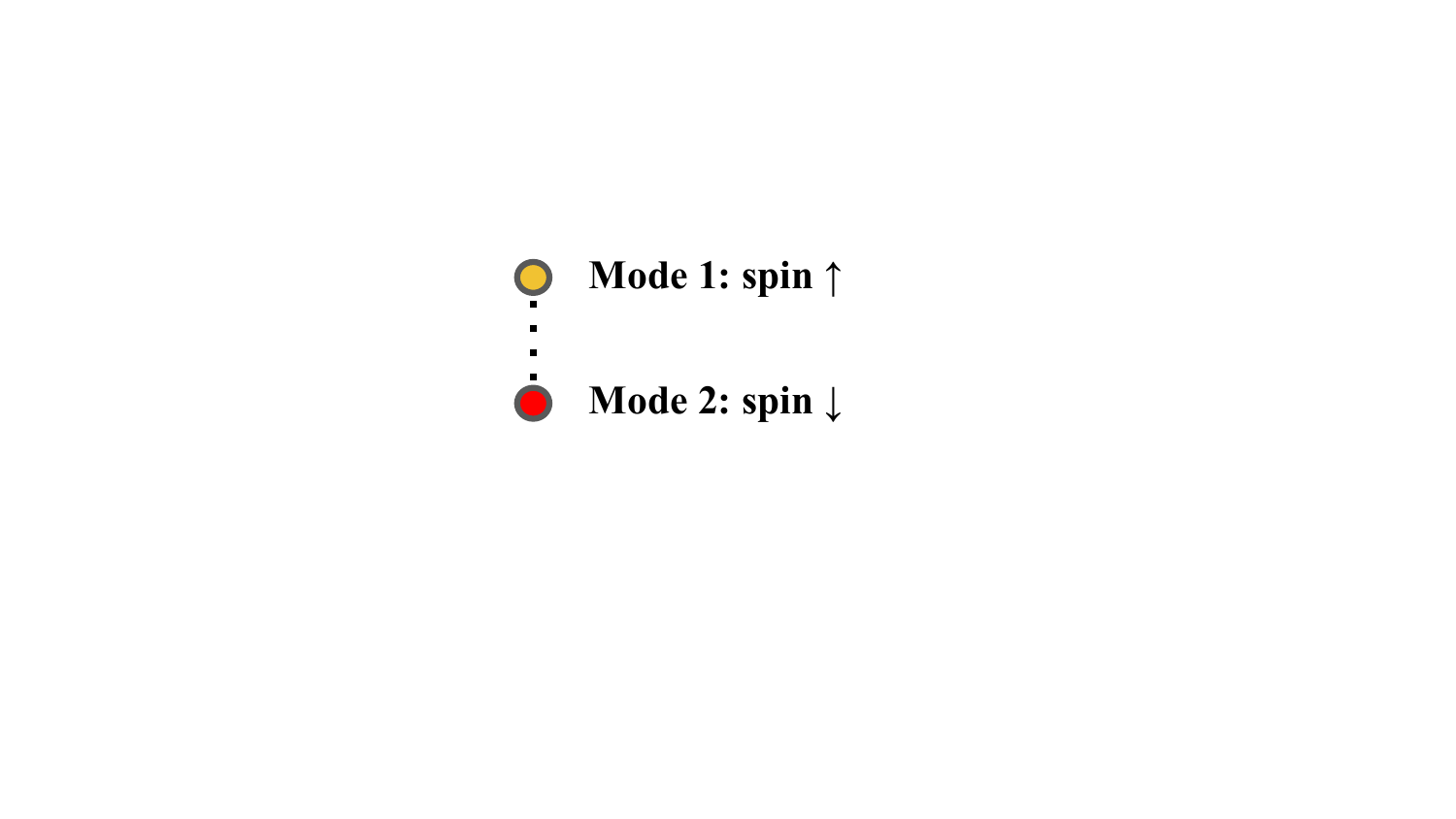}
    \caption{Diagrammatic representation of the single-site Hamiltonian in \eq{single_site_H}.}
    \label{fig:single_site_H}
\end{wrapfigure}
We begin by considering a single-site anharmonic oscillator Hamiltonian. The protocol to learn its parameters will be a core subroutine for the many-body case. Labelling the two spin modes of the single site as $1$ and $2$, as shown in \fig{single_site_H}, the single-site Hamiltonian is
\begin{equation}\label{eq:single_site_H}
    H = \omega_1n_1 + \omega_2n_2 + \xi_{12}n_1n_2.
\end{equation}
Note the slight change in notation relative to \eq{generic_H}: for simplicity, the subscripts 1 and 2 denote spin modes at the same spatial site, and there is no explicit spatial index. The only interaction term in this Hamiltonian is the on-site interaction $\xi_{12}n_1n_2$ between spin modes $1$ and $2$.

The parameters of this Hamiltonian can be learned using the robust phase estimation (RPE) algorithm of \citep{RPE_Kimmel_2015} (further analyzed and developed by \citep{RPE_Belliardo_2020, RPE_Russo_2021, RPE_Ni_2023}). RPE enables us to achieve Heisenberg scaling, and we provide a very brief review of the key ideas here. The goal is to estimate an unknown phase $\omega \in [-1,1]$, given the ability to apply a unitary of the form $U_{\omega} = e^{-i \omega H}$ (where $H$ is Hermitian) to a probe state $\ket{\psi_0}$ of our choosing and perform measurements on the resulting state. Depending on whether the metrological resource available is time or entanglement, either $U_{\omega}$ is applied multiple times to a single probe state (which can be viewed as applying the time evolution operator $e^{-i\omega H t}$ for some time $t$), or many copies of $U_{\omega}$ are simultaneously applied to multiple copies of multiple entangled probes. In this work we restrict our attention to the former case, where time is the resource. The RPE protocol requires performing multiple rounds of such experiments, with the rounds grouped into generations characterized by different amounts of evolution time. Suppose there are $M$ generations of experiments, where generation $j$ involves $m_j$ experiments, each of which involves time evolution $e^{-i\omega H t_j}$ for time $t_j$. The total time used for time evolution is $t_{\text{tot}} = \sum_1^M m_jt_j$. By the end of the process, we have an estimator $\tilde{\omega}$ of the true phase $\omega$, with some standard deviation $\Delta \tilde{\omega}$. The key point is that, as discussed in \citep{RPE_Kimmel_2015, RPE_Belliardo_2020}, the parameters $m_j$ and $t_j$ can be chosen such that $\Delta \tilde{\omega} \leq \mathcal{O}(1/t_{\text{tot}})$, which is the statement that RPE achieves Heisenberg scaling. More details on how to choose the specific values of these parameters can be found in \citep{RPE_Kimmel_2015, RPE_Belliardo_2020, RPE_Russo_2021, RPE_Ni_2023}.

In the approach to RPE of \citep{RPE_Kimmel_2015, RPE_Russo_2021, RPE_Belliardo_2020}, the measurement step in each experiment above is designed to output a Bernoulli random variable with a particular probability distribution. In particular, in each generation of experiments, one performs two types of experiments: ``Type-0" (with measurement outcomes labelled by 0 and 1) and ``Type-+" (with measurement outcomes $+$ or $-$). Their output probabilities in the $j^{\text{th}}$ generation of experiments are given by
\begin{equation}\label{eq:RPE_probs}
\begin{aligned}
    p_0(\omega, j) &= \frac{1}{2}(1+\cos(\omega t_j)) + \delta_0(j)\\
    p_+(\omega, j) &= \frac{1}{2}(1+\sin(\omega t_j)) + \delta_+(j)
\end{aligned}
\end{equation}
where $\delta_0, \delta_+$ represents additive noise, such as state-preparation-and-measurement (SPAM) error. As long as $\delta_0$ and $\delta_+$ are bounded by a small constant, the measurement outcomes obeying the above distribution can be post-processed to yield an estimate of $\omega$ with Heisenberg scaling. This is done by repeatedly halving the size of the confidence interval for $\omega$ using the outcomes of each successive generation of experiments. For a more detailed review of RPE we refer the reader to \citep{RPE_Kimmel_2015, RPE_Belliardo_2020, RPE_Russo_2021, RPE_Ni_2023}.

Returning to our problem of learning the parameters $\omega_1, \omega_2$, and $\xi_{12}$ of the Hamiltonian in $\eq{single_site_H}$, we use RPE to learn them one by one. In each case, we must design an experiment involving an appropriate initial state of the fermionic system, and choice of measurements, such that we obtain the desired Type-0 and Type-+ measurement statistics. We do this by preparing an initial state which, under time evolution, picks up a relative phase that depends in a known way on the Hamiltonian parameter to be learned, and measure the time-evolved state in bases that yield the desired statistics. 

Starting with $\omega_1$, we assume the availability of a fermionic ancilla spin mode, to which can couple a spin mode of our system. Let $b^{\dag}$ and $b$ respectively denote the creation and annihilation operators for the ancilla mode. To express our initial state and measurement basis, we define the following two-mode unitaries, which act on system mode $1$ and the ancilla mode:
\begin{equation}\label{eq:local_unitaries_def}
\begin{aligned}
    V(\theta) &= e^{\theta (\create{a}{1}\create{b}{} - \ann{b}{}\ann{a}{1})}\\
    W(\theta) &= e^{i \theta (\create{a}{1}\create{b}{} + \ann{b}{}\ann{a}{1})}
\end{aligned}
\end{equation}
For $\theta = -\frac{\pi}{4}$, $V$ and $W$ apply the following transformations to $\ket{\Omega}$, the vacuum state of the entire system (original system plus ancilla):
\begin{equation}\label{eq:local_unitaries_action}
\begin{aligned}
    V(-\tfrac{\pi}{4})\ket{\Omega} &= \frac{1}{\sqrt{2}}(\ket{\Omega} + \create{a}{1}\create{b}{}\ket{\Omega})\\
    W(-\tfrac{\pi}{4})\ket{\Omega} &= \frac{1}{\sqrt{2}}(\ket{\Omega} - i\create{a}{1}\create{b}{}\ket{\Omega})
\end{aligned}
\end{equation}
These are equally weighted superpositions of the vacuum state and the state with both modes occupied. They belong to the well-known class of fermionic Gaussian states \citep{bravyi2004lagrangian}.

Now, choose the initial state to be $V(-\tfrac{\pi}{4})\ket{\Omega}$. Next, induce dependence on the parameter $\omega_1$ by applying the time evolution operator $e^{-iHt}$, where $H$ is the Hamiltonian from \eq{single_site_H}:
\begin{equation}
    e^{-iHt}\bigl(V(-\tfrac{\pi}{4}\ket{\Omega})\bigr) = \frac{1}{\sqrt{2}}(\ket{\Omega} + e^{-i\omega_1 t}\create{a}{1}\create{b}{}\ket{\Omega})
\end{equation}
For the Type-0 measurement, apply $V^{\dag}(-\frac{\pi}{4})$, then measure the occupation numbers of system mode 1 and the ancilla mode. The probability of obtaining $\ket{\Omega}$ as the final state (corresponding to no occupation) is given by
\begin{equation}\label{eq:RPE_prob_0_omega}
    p_0 = |\bra{\Omega} V^{\dag}(-\tfrac{\pi}{4})\, e^{-iHt}\, V(-\tfrac{\pi}{4})\ket{\Omega} |^2 = \frac{1}{2}(1+\cos(\omega_1 t)).
\end{equation}
For the Type-+ measurement, apply $W^{\dag}(-\frac{\pi}{4})$, then measure the occupation numbers of system mode 1 and the ancilla mode. The probability of obtaining $\ket{\Omega}$ as the final state (interpreted as the ``+"-outcome due to the effective basis change implemented by $W^{\dag}$) is\\ 
\begin{equation}\label{eq:RPE_prob_+_omega}
    p_+ = |\bra{\Omega} W^{\dag}(-\tfrac{\pi}{4}) \, e^{-iHt}\, V(-\tfrac{\pi}{4}) \ket{\Omega} |^2 = \frac{1}{2}(1+\sin(\omega_1 t)).
\end{equation}
In practice, the probabilities above will be shifted by additive noise terms, as in \eq{RPE_probs}, corresponding to SPAM error (which, as discussed above, do not affect the RPE protocol as long as they are bounded by a small constant). With measurement statistics of this form, we can perform robust phase estimation as described above to learn $\omega_1$ with Heisenberg scaling. 

The discussion above also clarifies the utility of the ancilla. From a purely mathematical perspective, a more natural initial state to use would be $e^{\theta (a_1-a^{\dag}_1)}\ket{\Omega} = \frac{1}{\sqrt{2}}(\ket{\Omega} + \create{a}{1}\ket{\Omega})$. Then, replacing $V(\theta)$ by $e^{\theta (a_1-a^{\dag}_1)}$ and $W(\theta)$ by $e^{i \theta (a_1+a^{\dag}_1)}$ in the expressions above, would yield the same values for $p_0$ and $p_+$ as obtained in \eq{RPE_prob_0_omega} and \eq{RPE_prob_+_omega}. This approach seems simpler, as it does not require an ancilla, and the unitaries $V$ and $W$ act only on a single mode. However, the unitaries $V$ and $W$ are unphysical, because they can be used to prepare a superposition of an odd-parity and even-parity state, as we just saw. These states belong to an unphysical sector of our fermionic Hilbert space, because their existence would enable violations of the no-signalling principle \citep{Vidal_2021}. This \textit{parity superselection rule} forces us to use either odd parity or even parity states. If we prepared such states using just the system modes, the phases resulting from time evolution would involve linear combinations of the unknown coefficients. Therefore, to isolate $\omega_1$, we couple to the ancilla. However, this is not necessary for $\omega_2$, as we discuss below.

To learn the parameter $\omega_2$, there are two approaches: with or without the ancilla. The first option, which uses the ancilla, is to simply perform the same procedure described above, but interchanging the roles of modes $1$ and $2$. The second option is to use a different initial state that only involves the system (no ancilla), and takes advantage of the fact that we have already learned $\omega_1$. In particular, instead of using the $V$ and $W$ defined in \eq{local_unitaries_def}, we define new two-mode unitaries $V'$ and $W'$, which are fermionic beamsplitter operations acting on system modes $1$ and $2$:
\begin{equation}\label{eq:local_unitaries_def_2}
\begin{aligned}
    V'(\theta) &= e^{\theta (\ygen{a}{1}{2})}\\
    W'(\theta) &= e^{i \theta (\xgen{a}{1}{2})}
\end{aligned}
\end{equation}
For $\theta = -\frac{\pi}{4}$, $V'$ and $W'$ apply the following transformations to the initial state with only mode $1$ occupied, $\create{a}{1}\ket{\Omega}$:
\begin{equation}\label{eq:local_unitaries_action_2}
\begin{aligned}
    V'(-\tfrac{\pi}{4})(\create{a}{1}\ket{\Omega}) &= \frac{1}{\sqrt{2}}(\create{a}{1}\ket{\Omega} + \create{a}{2}\ket{\Omega})\\
    W'(-\tfrac{\pi}{4})(\create{a}{1}\ket{\Omega}) &= \frac{1}{\sqrt{2}}(\create{a}{1}\ket{\Omega} - i\create{a}{2}\ket{\Omega})
\end{aligned}
\end{equation}
The initial and final states in \eq{local_unitaries_action_2} are both odd-parity states, so these states and operations satisfy the parity superselection rule. (Note that having odd parity, these are not Gaussian states, as the latter type of states are always of even parity. Of course, Theorem \ref{main_thm} still stands, namely that the ability to prepare Gaussian states is sufficient to learn the parameters $-$ if one only has the ability to prepare Gaussian states, $\omega_2$ can be learned by the technique used for $\omega_1$. However, if one has the ability to prepare the specified odd-parity states, an ancilla need not be used to learn $\omega_2$.) Now, the parameter-dependence induced by time evolving the state $V'(-\tfrac{\pi}{4})(\create{a}{1}\ket{\Omega})$ under $e^{-iHt}$ is
\begin{equation}
    e^{-iHt}V'(-\tfrac{\pi}{4})(\create{a}{1}\ket{\Omega}) = \frac{1}{\sqrt{2}}(e^{-i\omega_1 t}\create{a}{1}\ket{\Omega} + e^{-i\omega_2 t}\create{a}{2}\ket{\Omega}) = \frac{1}{\sqrt{2}}e^{-i\omega_1 t}(\create{a}{1}\ket{\Omega} + e^{-i(\omega_2-\omega_1)t}\create{a}{2}\ket{\Omega}).
\end{equation}
For the Type-0 measurement, apply $V'^{\dag}(-\frac{\pi}{4})$ after time evolution, then measure the occupation numbers of modes $1$ and $2$. The probability of obtaining the state $\create{a}{1}\ket{\Omega}$ upon measurement (mode $1$ occupied, mode $2$ unoccupied) is given by
\begin{equation}\label{eq:RPE_prob_0_omega_2}
    p_0 = |(\bra{\Omega} \ann{a}{1}) V'^{\dag}(-\tfrac{\pi}{4})\, e^{-iHt}\, V'(-\tfrac{\pi}{4})(\create{a}{1}\ket{\Omega}) |^2 = \frac{1}{2}\bigl(1+\cos((\omega_2-\omega_1) t)\bigr).
\end{equation}
For the Type-+ measurement, apply $W'^{\dag}(-\frac{\pi}{4})$ after time evolution, then measure the occupation numbers of modes $1$ and $2$. The probability of obtaining $\create{a}{1}\ket{\Omega}$ upon measurement is\\ 
\begin{equation}\label{eq:RPE_prob_+_omega_2}
    p_+ = |(\bra{\Omega} \ann{a}{1}) W'^{\dag}(-\tfrac{\pi}{4}) \, e^{-iHt}\, V'(-\tfrac{\pi}{4}) (\create{a}{1}\ket{\Omega}) |^2 = \frac{1}{2}\bigl(1+\sin((\omega_2-\omega_1) t)\bigr).
\end{equation}
Applying RPE to the measurement statistics, as described previously, provides an estimate of ($\omega_2-\omega_1)$ with Heisenberg scaling. Adding this to the previously estimated value of $\omega_1$ provides an estimate of $\omega_2$. Appendix \ref{appendix:app_b} discusses the RMS error of such linear combinations of estimators.

Finally, to learn the last coupling coefficient $\xi_{12}$, no ancilla is needed. The procedure is almost the same as that for $\omega_1$, the only difference being that the unitaries $V(-\tfrac{\pi}{4})$ and $W(-\tfrac{\pi}{4})$ defined in \eq{local_unitaries_def} now act on the two system modes (that is, $b$ in \eq{local_unitaries_def} is replaced by $a_2$). The Type-0 and Type-+ measurement probabilities are:
\begin{equation}\label{eq:RPE_probs_xi}
\begin{aligned}
    p_0 &= |\bra{\Omega}V^{\dag}(-\tfrac{\pi}{4}) \, e^{-iHt}\, V(-\tfrac{\pi}{4})\ket{\Omega} |^2 = \frac{1}{2}\bigl(1+\cos((\omega_1 + \omega_2 + \xi_{12})t)\bigr)\\
    p_+ &= |\bra{\Omega}W^{\dag}(-\tfrac{\pi}{4}) \, e^{-iHt}\, V(-\tfrac{\pi}{4})\ket{\Omega} |^2 = \frac{1}{2}\bigl(1+\sin((\omega_1 + \omega_2 + \xi_{12})t)\bigr)
\end{aligned}
\end{equation}
Performing RPE with this data provides an estimate of the sum ($\omega_1 + \omega_2 + \xi_{12}$), which can be used in conjunction with the previously obtained estimates of $\omega_1$ and $\omega_2$ to obtain an estimate of $\xi_{12}$. In this way, all three unknown parameters of the single-site Hamiltonian in \eq{single_site_H} can be learned. 

\subsection{Learning a two-site (four-mode) Hamiltonian}\label{sec:two_coupled_oscillators}
\begin{wrapfigure}{r}{0.25\textwidth}
    \centering
    \includegraphics[width=0.25\textwidth]{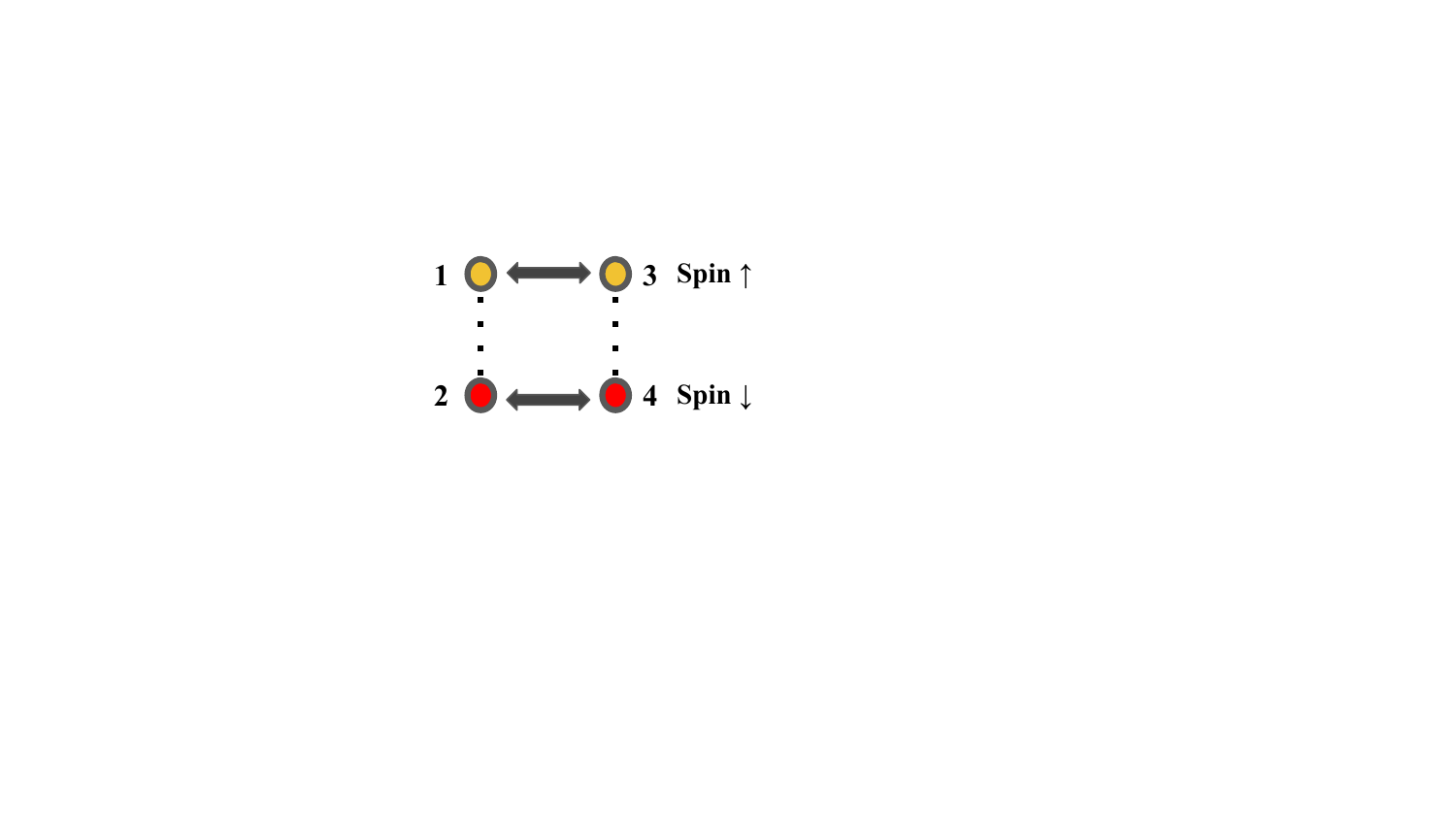}
    \caption{Diagrammatic representation of the two-site Hamiltonian in \eq{two_site_H}.}
    \label{fig:two_site_H}
\end{wrapfigure}
Next, consider a two-site Hamiltonian representing two coupled fermionic quantum anharmonic oscillators. Opposite-spin modes $1$ and $2$ constitute one spatial site, and opposite-spin modes $3$ and $4$ constitute the other, as shown in \fig{two_site_H}. The spin modes at a given site are coupled by an interaction term, as in \sect{single_site_oscillator}, while modes of the same spin in adjacent sites are coupled by a hopping interaction. Thus the Hamiltonian takes the following form, where all indices refer to spin modes of the system:
\begin{equation}\label{eq:two_site_H}
\begin{aligned}
    H &= h_{13}\create{a}{1}\ann{a}{3} + h_{31}\create{a}{3}\ann{a}{1} + h_{24}\create{a}{2}\ann{a}{4} + h_{42}\create{a}{4}\ann{a}{2}\\
    &+ \omega_1n_1 + \omega_2n_2 + \omega_3n_3 + \omega_4n_4 + \xi_{12}n_1n_2 + \xi_{34}n_3n_4
\end{aligned}
\end{equation}

We can learn this Hamiltonian in two stages. First, we learn the coefficients of the number-conserving terms, i.e. the $\omega$ (chemical potential) and $\xi$ (on-site interaction) coefficients in the second line of \eq{two_site_H}. This involves `discrete quantum control' $-$ reshaping the Hamiltonian using random unitaries that decouple the two spatial sites by approximately eliminating the hopping terms (as in refs \citep{Huang_qubit_2023, Li_boson_2023}). Next, we learn the hopping coefficients in the first line of \eq{two_site_H}, which involves a performing a basis transformation and using the reshaping technique again. We elaborate on the details of this procedure below.

\subsubsection{Learning coefficients of number-conserving terms}\label{sec:number_conserving_coeffs}
To learn the $\omega$ coefficients, we require one ancilla mode per spatial site, that is, one for system modes $1$ and $2$, and another for system modes $3$ and $4$. These ancilla modes are first used to learn $\omega_1$ and $\omega_3$ in parallel. They can optionally be re-used to learn $\omega_2$ and $\omega_4$ in parallel, although those coefficients can be learned without ancillae as well. Finally, $\xi_{12}$ and $\xi_{34}$ are learned in parallel, without ancillae. 

Starting with coefficients $\omega_1$ and $\omega_3$, the strategy is to alternate small time-steps of Hamiltonian evolution generated by $H$ with random unitaries sampled from a suitable distribution. The resulting time evolution is then approximately generated by a `reshaped' effective Hamiltonian $H_{\text{eff}}$:
\begin{equation}\label{eq:H_eff_step1}
    H_{\text{eff}} = \omega_1n_1 + \omega_2n_2 + \omega_3n_3 + \omega_4n_4 + \xi_{12}n_1n_2 + \xi_{34}n_3n_4
\end{equation}

Techniques to perform this reshaping, inspired by dynamical decoupling and the qDRIFT randomized compiling algorithm \citep{Campbell_2019}, were developed in \citep{Huang_qubit_2023} for low-intersection qubit Hamiltonians and in \citep{Li_boson_2023} for a class of Bose-Hubbard Hamiltonians. The latter class of Hamiltonians \citep{Li_boson_2023} is similar to ours. Here we show how this approach extends to the fermionic case, using fermionic analogues of the unitaries used in the bosonic case. 

To perform reshaping, we begin by defining a distribution $\mathcal{D}$ over unitaries $U$ such that:
\begin{equation}
    H_{\text{eff}} = \mathbb{E}_{U \sim \mathcal{D}} U^{\dag} H U
\end{equation}
where $\tsub{H}{eff}$ has the form shown in \eq{H_eff_step1}. In particular, the distribution that we use is defined over the following one-parameter subset of `fermionic linear optics' (FLO) \citep{bravyi2004lagrangian} unitaries:
\begin{equation}\label{eq:reshaping_dist}
    U = e^{-i\theta (n_1 + n_2)}
\end{equation}
The distribution $\mathcal{D}$ corresponds to $\theta \sim \mathcal{U}([0,2\pi])$, the uniform distribution over the interval $[0,2\pi]$. Using the following facts that hold for any fermionic mode $j$ and any pair of distinct modes $j \neq k$: 
\begin{equation}\label{eq:fermionic_identities}
\begin{aligned}
     e^{i\theta n_j} \create{a}{j} e^{-i\theta n_j} = e^{i\theta} \create{a}{j}\\
     [a_j,n_k] = 0
\end{aligned}  
\end{equation}
the distribution defined by \eq{reshaping_dist} performs the desired reshaping:
\begin{equation}\label{eq:reshaping_integral}
\begin{aligned}
     H_{\text{eff}} = \mathbb{E}_{U \sim \mathcal{D}} U^{\dag} H U &= \frac{1}{2\pi}\int_{0}^{2\pi} d\theta  e^{i\theta (n_1 + n_2)} H e^{-i\theta (n_1 + n_2)}\\
     &= \omega_1n_1 + \omega_2n_2 + \omega_3n_3 + \omega_4n_4 + \xi_{12}n_1n_2 + \xi_{34}n_3n_4
\end{aligned}  
\end{equation}
As noted in \citep{Li_boson_2023} for the bosonic setting, the above reshaping can be interpreted as enforcing the U(1) symmetry of particle number conservation on the first spatial site of the system, comprising spin modes 1 and 2.

Notice that system modes $1$ and $2$ evolve independently from modes $3$ and $4$ in the dynamics generated by $H_{\text{eff}}$ because no terms couple the two pairs. Therefore, if we could perform time evolution $\exp(-iH_{\text{eff}}t)$ generated by the reshaped Hamiltonian $H_{\text{eff}}$ rather than the original Hamiltonian $H$, we could use robust phase estimation as described in \sect{single_site_oscillator} to learn the coefficients corresponding to the two sites in parallel, due to their decoupling. In practice, we will not be able to perform $\exp(-iH_{\text{eff}}t)$ exactly, but can approximate it by the following sequence of operators:
\begin{equation}\label{eq:actual_op}
    e^{-iH_{\text{eff}}t} \approx \prod_{j=1}^r U_j^{\dag} e^{-iHt/r} U_j
\end{equation}
where each $U_j$ is independently sampled from the distribution $\mathcal{D}$. In the limit $r \to \infty$, the above time evolution approaches the desired time evolution $\exp(-iH_{\text{eff}}t)$. In Appendix \ref{appendix:app_a}, we discuss the following bound that holds for the expectation values of any operator $O$ with finite ($\order{1}$) support and $\norm{O} \leq 1$:
\begin{equation}\label{eq:operator_error_bound}
    \left| \text{Tr}(\rho(t)_{\text{approx}} O) - \text{Tr}(\rho(t)_{\text{exact}} O)\right| \leq \mathcal{O}\left(\frac{t^2 \lambda^2_{max}}{r}\right)
\end{equation}
where
\begin{equation}
    \rho(t)_{\text{approx}} = \mathbb{E}_{U_j \sim \mathcal{D}}\biggl(\displaystyle{\prod_{1\leq j \leq r}^{\leftarrow}} U_j^{\dag} e^{-iHt/r} U_j\biggr) \rho \biggl(\prod_{1\leq j \leq r}^{\rightarrow} U_j^{\dag} e^{iHt/r} U_j\biggr),
\end{equation}
$\rho(t)_{\text{exact}} = e^{-iH_{\text{eff}}t} \rho e^{iH_{\text{eff}}t}$, and $\lambda_{max}$ is the largest absolute value among the Hamiltonian coefficients, which we henceforth take to be $1$.

The outcome probabilities $p_0$ and $p_+$ of the RPE experiments in \sect{single_site_oscillator} (\eq{RPE_prob_0_omega}, \eq{RPE_prob_+_omega} and \eq{RPE_probs_xi}) can manifestly be expressed as expectation values of projection operators, which have norm $1$. So by \eq{operator_error_bound}, performing RPE with the approximate time evolution operator in \eq{actual_op} shifts the output probabilities by an additive error of magnitude $\mathcal{O}(\frac{t^2}{r})$. By choosing $r$ such that this term is $\mathcal{O}(1)$ and sufficiently small, this additive error can be swept into the noise terms $\delta_0, \delta_+$ (c.f. \eq{RPE_probs}), and RPE can still be accurately performed. 

Thus, performing RPE with the operator in \eq{actual_op}, via the protocol described in \sect{single_site_oscillator}, on the decoupled subsystems in parallel, allows us to learn $\omega_1$ and $\omega_3$ simultaneously with Heisenberg scaling. In fact, the procedure described above can be slightly simplified to achieve the same result. Currently, the unitaries $e^{-i\theta (n_1 + n_2)}$ used for reshaping eliminate both sets of hopping terms in the original Hamiltonian of \eq{two_site_H}. This has the (approximate) effect of decoupling the first spatial site (modes 1 and 2) \textit{completely} from the second spatial site (modes 3 and 4). This complete decoupling implies that we could equally well have chosen to learn $\omega_4$, rather than $\omega_3$, in parallel with $\omega_1$, by changing the ancilla coupling accordingly. But if we choose specifically to learn $\omega_1$ and $\omega_3$ in parallel, partial decoupling would suffice. In particular, instead of using the two-mode unitaries $e^{-i\theta (n_1 + n_2)}$ for reshaping, we could use the single-mode unitaries $e^{-i\theta n_1}$ that only act on mode $1$. This would eliminate the hopping terms $h_{13}\create{a}{1}\ann{a}{3} + h_{31}\create{a}{3}\ann{a}{1}$ but not $h_{24}\create{a}{2}\ann{a}{4} + h_{42}\create{a}{4}\ann{a}{2}$. Due to our choice of initial states described in \sect{single_site_oscillator}, the latter pair of hopping terms would have no physical effect, since modes $2$ and $4$ would remain unoccupied throughout the experiment.

Next, to learn $\omega_2$ and $\omega_4$, there are two approaches, as described in \sect{single_site_oscillator} for the single-site case. The approach with ancillae is analogous to the approach described above to learn $\omega_1$ and $\omega_3$, in which case partial decoupling would suffice (that is, $e^{-i\theta n_2}$ can be used for reshaping, allowing $h_{13}\create{a}{1}\ann{a}{3} + h_{31}\create{a}{3}\ann{a}{1}$ to remain in the Hamiltonian). On the other hand, in the ancilla-free approach, full decoupling is required -- both sets of hopping terms must be effectively eliminated. If $h_{13}\create{a}{1}\ann{a}{3} + h_{31}\create{a}{3}\ann{a}{1}$ remains in the reshaped Hamiltonian, this acts nontrivially on the states of the form in \eq{local_unitaries_action_2}. The ancilla-free procedure described in \sect{single_site_oscillator} then no longer yields the desired measurement statistics. Hence, the ancilla-free approach requires reshaping with $e^{-i\theta (n_1 + n_2)}$. To learn the coefficients $\omega_2$ and $\omega_4$, we therefore have a tradeoff: the ancilla-based approach allows reshaping with single-mode unitaries at the expense of ancillary modes, while the ancilla-free approach requires two-mode unitaries for reshaping but no ancillary resources.

Finally, to learn $\xi_{12}$ and $\xi_{34}$, the procedure is almost identical to that used for $\omega_1$ and $\omega_3$, except that ancillae are not involved. As in the single-site case of \sect{single_site_oscillator}, the initial states are modified accordingly.

\subsubsection{Learning coefficients of hopping terms}\label{sec:learn_hopping_coeffs}
To learn the (generically complex) hopping amplitudes, $h_{13}$ and $h_{24}$, we offer two approaches, one ancilla-dependent and the other ancilla-free. In both cases, we first perform a change of basis via a Bogoliubov transformation. In the new basis, the real or imaginary components of $h_{13}$ and $h_{24}$ appear as coefficients of number-conserving terms, enabling us to learn these parameters using the reshaping approach as in \sect{number_conserving_coeffs} above. We will need to use different basis transformations to learn the real and imaginary components.
The basis transformations on modes $1$ and $3$ are defined by the two-mode FLO unitaries $U_x^{(1,3)}$ and $U_y^{(1,3)}$ (representing fermionic beamsplitter operations as in \eq{local_unitaries_def_2}):
\begin{equation}\label{eq:basis_unitaries}
\begin{aligned}
     U_x^{(1,3)}(\theta) &= e^{i \theta (\xgen{a}{1}{3})} = \mathbb{I} + (\cos\theta - 1)(n_1 - n_3)^2 + i\sin\theta (\xgen{a}{1}{3})\\
     U_y^{(1,3)}(\theta) &= e^{\theta (\ygen{a}{1}{3})} = \mathbb{I} + (\cos\theta - 1)(n_1 - n_3)^2 - \sin\theta (\ygen{a}{1}{3}).
\end{aligned}  
\end{equation}
The annihilation operators $a_1$ and $a_3$ are transformed as follows:
\begin{equation}\label{eq:transform_unitaries}
\begin{aligned}
    \begin{pmatrix}
         U_x^{(1,3)}(\theta)\, a_1\, U_x^{(1,3) \dag}(\theta)\\
          U_x^{(1,3)}(\theta) \,a_3\, U_x^{(1,3) \dag}(\theta)
     \end{pmatrix}
     &= \begin{pmatrix}
         \cos\theta & -i\sin\theta\\
         -i\sin\theta & \cos\theta
     \end{pmatrix}
     \begin{pmatrix}
         a_1\\
         a_3
     \end{pmatrix}\\
     \begin{pmatrix}
         U_y^{(1,3)}(\theta)\, a_1\, U_y^{(1,3) \dag}(\theta)\\
          U_y^{(1,3)}(\theta) \,a_3\, U_y^{(1,3) \dag}(\theta)
     \end{pmatrix}
     &= \begin{pmatrix}
         \cos\theta & \sin\theta\\
         -\sin\theta & \cos\theta
     \end{pmatrix}
     \begin{pmatrix}
         a_1\\
         a_3
     \end{pmatrix}.
\end{aligned}
\end{equation}
The basis transformations on modes $2$ and $4$ are given by the analogously defined unitaries $U_x^{(2,4)}$ and $U_y^{(2,4)}$. Comparing to the case of bosonic operators explored in \citep{Li_boson_2023}, the $U_x$-transformation differs by a minus sign on the off-diagonal terms, while the $U_y$-transformation is identical.

In both the ancilla-based and the ancilla-free protocols, the hopping coefficients are learned in two steps. First, we use the $U_y$ basis to learn the real components, Re($h_{13}$) and Re($h_{24}$); then we use the $U_x$ basis to learn the imaginary components, Im($h_{13}$) and Im($h_{24}$).

Starting with the real components, consider the transformation obtained by simultaneously rotating all four modes using $U_y^{(1,3)}(\theta)$ and $U_y^{(2,4)}(\theta)$ at $\theta = \frac{\pi}{4}$. The transformed fermionic operators $\tilde{a}_1, \tilde{a}_2, \tilde{a}_3, \tilde{a}_4$ are given by
\begin{equation}\label{eq:mode_transfs_Uy}
\begin{aligned}
    \begin{pmatrix}
         \tilde{a}_1 \\
         \tilde{a}_3
     \end{pmatrix}
    &= 
    \begin{pmatrix}
         U_y^{(1,3)}(\tfrac{\pi}{4})\, a_1\, U_y^{\dag (1,3)}(\tfrac{\pi}{4})\\
         U_y^{(1,3)}(\tfrac{\pi}{4}) \,a_3\, U_y^{\dag (1,3)}(\tfrac{\pi}{4})
     \end{pmatrix}
     = \frac{1}{\sqrt{2}}\begin{pmatrix}
         1 & 1\\
         -1 & 1
     \end{pmatrix}
     \begin{pmatrix}
         a_1\\
         a_3
     \end{pmatrix}\\
     \begin{pmatrix}
         \tilde{a}_2 \\
         \tilde{a}_4
     \end{pmatrix}
    &= 
    \begin{pmatrix}
         U_y^{(2,4)}(\tfrac{\pi}{4})\, a_2\, U_y^{\dag (2,4)}(\tfrac{\pi}{4})\\
         U_y^{(2,4)}\tfrac{\pi}{4}) \,a_4\, U_y^{\dag (2,4)}(\tfrac{\pi}{4})
     \end{pmatrix}
     = \frac{1}{\sqrt{2}}\begin{pmatrix}
         1 & 1\\
         -1 & 1
     \end{pmatrix}
     \begin{pmatrix}
         a_2\\
         a_4
     \end{pmatrix}.
\end{aligned}
\end{equation}
Under this basis change, our two-site Hamiltonian in \eq{two_site_H} is transformed as follows:
\begin{equation}\label{eq:two_site_H_Uy}
\begin{aligned}
    H \mapsto \tilde{H} &= \biggl(\frac{\omega_1+\omega_3}{2} + \text{Re}(h_{13})\biggr)\tilde{n}_1 + \biggl(\frac{\omega_2+\omega_4}{2} + \text{Re}(h_{24})\biggr)\tilde{n}_2\\
    &+ \biggl(\frac{\omega_1+\omega_3}{2} - \text{Re}(h_{13})\biggr)\tilde{n}_3 + \biggl(\frac{\omega_2+\omega_4}{2} - \text{Re}(h_{24})\biggr)\tilde{n}_4\\
    &+ \biggl(\frac{\xi_{12}+\xi_{34}}{4}\biggr)(\tilde{n}_1\tilde{n}_2 + \tilde{n}_2\tilde{n}_3 + \tilde{n}_3\tilde{n}_4 + \tilde{n}_1\tilde{n}_4) \\
    &+ \biggl(\frac{\omega_3-\omega_1}{2}\biggr)(\xgen{\tilde{a}}{1}{3}) + \biggl(\frac{\omega_4-\omega_2}{2}\biggr)(\xgen{\tilde{a}}{2}{4}) \\
    &+ \biggl(\frac{\xi_{34}-\xi_{12}}{4}\biggr)\bigl((\tilde{n}_1 + \tilde{n}_3)(\xgen{\tilde{a}}{2}{4}) + (\tilde{n}_2 + \tilde{n}_4)(\xgen{\tilde{a}}{1}{3})\bigr)\\
    &+ \biggl(\frac{\xi_{12}+\xi_{34}}{4}\biggr)(\xgen{\tilde{a}}{1}{3})(\xgen{\tilde{a}}{2}{4}).
\end{aligned}
\end{equation}
It is evident from this expression that learning the coefficients of $\tilde{n}_1$ and $\tilde{n}_2$ in $\tilde{H}$ (that is, the coefficients of the first two terms in \eq{two_site_H_Uy}) is sufficient to learn Re($h_{13}$) and Re($h_{24}$), assuming $\omega_1, \omega_2, \omega_3$ and $\omega_4$ have already been learned using the technique of \sect{number_conserving_coeffs}. 

To learn the desired parameters of $\tilde{H}$, we will apply the reshaping technique as in \sect{number_conserving_coeffs}, but in the $U_y$-basis. In this case, the distribution $\mathcal{D}$ used for reshaping is defined by unitaries of the form $e^{-i\theta (\tilde{n}_1 + \tilde{n}_2)}$ with $\theta \sim \mathcal{U}([0,2\pi])$. Note that these number-preserving operators in the $U_y$-basis can be expressed in terms of the $U_x$ operators introduced in \eq{basis_unitaries},\\
\begin{equation}\label{eq:number_Ux_equiv}
\begin{aligned}
     e^{i\theta\tilde{n}_1} &= e^{i\frac{\theta}{2}(n_1+n_3)} U_x^{(1,3)}(\tfrac{\theta}{2})\\ 
     e^{i\theta\tilde{n}_2} &= e^{i\frac{\theta}{2}(n_2+n_4)} U_x^{(2,4)}(\tfrac{\theta}{2}).
\end{aligned}  
\end{equation}
So, defining $U_x (\theta) = U_x^{(1,3)}(\theta)U_x^{(2,4)}(\theta)$, we have $e^{i\theta(\tilde{n}_1 + \tilde{n}_2)} = e^{i\frac{\theta}{2}(n_1+n_2+n_3+n_4)} U_x(\tfrac{\theta}{2})$. Since the total number of particles across all four modes, $n_1 + n_2 + n_3 + n_4$, is conserved by the Hamiltonian $H$, conjugation by $e^{-i\theta (\tilde{n}_1 + \tilde{n}_2)}$ is equivalent to conjugation by $U_x (\theta)$:
\begin{equation}
     e^{i\theta (\tilde{n}_1 + \tilde{n}_2)} H e^{-i\theta (\tilde{n}_1 + \tilde{n}_2)} = U_x(\tfrac{\theta}{2}) H U_x(-\tfrac{\theta}{2}).
\end{equation}
So averaging over the distribution $\mathcal{D}$ is equivalent to averaging over $U_x(\tfrac{\theta}{2})$ with $\theta \sim \mathcal{U}([0,2\pi])$. 

From the properties of fermionic operators listed in \eq{fermionic_identities}, it follows that conjugation of $\tilde{H}$ by $U \sim \mathcal{D}$ cancels, in expectation, the terms that are not particle-number conserving in either of the first two modes (in the $U_y$-basis). This leaves the following reshaped Hamiltonian:
\begin{equation}\label{eq:reshaping_integral_Uy}
\begin{aligned}
     \tilde{H}_{\text{eff}} = \mathbb{E}_{U \sim \mathcal{D}} U^{\dag} \tilde{H} U &= \frac{1}{2\pi}\int_{0}^{2\pi} d\theta \,  e^{i\theta (\tilde{n_1} + \tilde{n_2})} H e^{-i\theta (\tilde{n_1} + \tilde{n_2})} = \frac{1}{2\pi}\int_{0}^{2\pi} d\theta \, U_x(\tfrac{\theta}{2}) H U_x(-\tfrac{\theta}{2})\\
    &= \biggl(\frac{\omega_1+\omega_3}{2} + \text{Re}(h_{13})\biggr)\tilde{n}_1 + \biggl(\frac{\omega_2+\omega_4}{2} + \text{Re}(h_{24})\biggr)\tilde{n}_2\\
    &+ \biggl(\frac{\omega_1+\omega_3}{2} - \text{Re}(h_{13})\biggr)\tilde{n}_3 + \biggl(\frac{\omega_2+\omega_4}{2} - \text{Re}(h_{24})\biggr)\tilde{n}_4\\
    &+ \biggl(\frac{\xi_{12}+\xi_{34}}{4}\biggr)(\tilde{n}_1\tilde{n}_2 + \tilde{n}_2\tilde{n}_3 + \tilde{n}_3\tilde{n}_4 + \tilde{n}_1\tilde{n}_4).
\end{aligned}  
\end{equation}

As in \sect{number_conserving_coeffs}, in practice we will use an approximate version of the time evolution operator corresponding to the reshaped Hamiltonian, obtained by sampling a finite number of unitaries from $\mathcal{D}$:
\begin{equation}\label{eq:actual_op_Uy}
    \prod_{j=1}^r U_x\biggl(\frac{\theta_j}{2}\biggr) e^{-iHt/r} U_x\biggl(-\frac{\theta_j}{2}\biggr)
\end{equation}

Now, to perform RPE, we can proceed either with an ancilla or without.

\subsubsection*{a. Learning hopping coefficients: ancilla-based protocol}

A single ancilla is used to learn the coefficients of $\tilde{n}_1$ and $\tilde{n}_2$ in turn. To specify the RPE protocol, we need analogues of the states $V(-\frac{\pi}{4})\ket{\Omega}$ and $W(-\frac{\pi}{4})\ket{\Omega}$ that appeared in the single-site protocol (\eq{local_unitaries_action} of \sect{single_site_oscillator}), in the $U_y$-basis. Using $b^{\dag}$ to denote the creation operator for the ancilla mode, the required states are given by
\begin{equation}\label{eq:rotated_probe_states}
\begin{aligned}
    U_y^{(1,3)}(\tfrac{\pi}{4})V(-\tfrac{\pi}{4})\ket{\Omega} &= \frac{1}{\sqrt{2}}(\ket{\Omega} + \create{\tilde{a}}{1}\create{b}{}\ket{\Omega})\\
    U_y^{(1,3)}(\tfrac{\pi}{4})W(-\tfrac{\pi}{4})\ket{\Omega} &= \frac{1}{\sqrt{2}}(\ket{\Omega} - i\create{\tilde{a}}{1}\create{b}{}\ket{\Omega})
\end{aligned}
\end{equation}
where $V(-\frac{\pi}{4})\ket{\Omega}$ and $W(-\frac{\pi}{4})\ket{\Omega}$ are the same two-mode operators introduced in \eq{local_unitaries_def}, which act on system mode $1$ (in the original basis) and the ancilla mode. \eq{rotated_probe_states} can be easily derived using the properties $\create{\tilde{a}}{1}U_y^{(1,3)}(\tfrac{\pi}{4}) = U_y^{(1,3)}(\tfrac{\pi}{4})\create{a}{1}$ and $U_y^{(1,3)}(\tfrac{\pi}{4})\ket{\Omega} = \ket{\Omega}$.

Having defined these states, the rest of the procedure is completely analogous to that used to learn $\omega_1$ in \sect{single_site_oscillator}, with time evolution being generated by the approximate operator in \eq{actual_op_Uy}. For the Type-0 experiment, apply the following sequence of operators to the initial state:
\begin{equation}
   V^{\dag}(-\tfrac{\pi}{4})U_y^{\dag (1,3)}(\tfrac{\pi}{4})\biggl(\prod_{j=1}^r U_x\biggl(\frac{\theta_j}{2}\biggr) e^{-iHt/r} U_x\biggl(-\frac{\theta_j}{2}\biggr)\biggr) U_y^{(1,3)}(\tfrac{\pi}{4})V(-\tfrac{\pi}{4})\ket{\Omega}.
\end{equation}
After measuring the occupation of system mode $1$ and the ancilla, the probability of returning to the vacuum state is given by
\begin{equation}
\begin{aligned}
   p_0 &= |\bra{\Omega}V^{\dag}(-\tfrac{\pi}{4})U_y^{\dag (1,3)}(\tfrac{\pi}{4})\biggl(\prod_{j=1}^r U_x\biggl(\frac{\theta_j}{2}\biggr) e^{-iHt/r} U_x\biggl(-\frac{\theta_j}{2}\biggr)\biggr) U_y^{(1,3)}(\tfrac{\pi}{4})V(-\tfrac{\pi}{4})\ket{\Omega}|^2\\  
   &= \frac{1}{2}\biggl(1+\cos\bigl(\frac{\omega_1+\omega_3}{2} + \text{Re}(h_{13})\bigr)\biggr) + \delta_0
\end{aligned}
\end{equation}
where the constant additive error $\delta_0$ comprises the approximation error in the time evolution operator and SPAM error.
For the Type-+ experiment, apply a similar sequence of operators to the initial state, but with $V^{\dag}(-\tfrac{\pi}{4})$ replaced by $W^{\dag}(-\tfrac{\pi}{4})$:
\begin{equation}
   W^{\dag}(-\tfrac{\pi}{4})U_y^{\dag (1,3)}(\tfrac{\pi}{4})\biggl(\prod_{j=1}^r U_x\biggl(\frac{\theta_j}{2}\biggr) e^{-iHt/r} U_x\biggl(-\frac{\theta_j}{2}\biggr)\biggr) U_y^{(1,3)}(\tfrac{\pi}{4})V(-\tfrac{\pi}{4})\ket{\Omega}
\end{equation}
Then the probability of returning to the vacuum state upon measurement is given by
\begin{equation}
\begin{aligned}
   p_+ &= |\bra{\Omega}W^{\dag}(-\tfrac{\pi}{4})U_y^{\dag (1,3)}(\tfrac{\pi}{4})\biggl(\prod_{j=1}^r U_x\biggl(\frac{\theta_j}{2}\biggr) e^{-iHt/r} U_x\biggl(-\frac{\theta_j}{2}\biggr)\biggr) U_y^{(1,3)}(\tfrac{\pi}{4})V(-\tfrac{\pi}{4})\ket{\Omega}|^2\\ 
   &= \frac{1}{2}\biggl(1+\sin\bigl(\frac{\omega_1+\omega_3}{2} + \text{Re}(h_{13})\bigr)\biggr) + \delta_+.
\end{aligned}
\end{equation}

As before, these experiments are run for varying amounts of time evolution and repetitions in accordance with the RPE procedure, to generate an estimate of $\frac{\omega_1+\omega_3}{2} + \text{Re}(h_{13})$ with Heisenberg scaling. Using the previously obtained estimates of $\omega_1$ and $\omega_3$, this yields an estimate of Re($h_{13}$). Appendix \ref{appendix:app_b} justifies the use of linear combinations of estimators in this way, even though the estimates of $\omega_1$ and $\omega_3$ are not truly independent (having been learned in parallel with approximate, rather than exact, decoupling of modes $1$ and $3$).

As noted in the single site case of \sect{single_site_oscillator}, one possible benefit of the ancilla-based protocol over the ancilla-free one is that, due to the mode occupancies of the initial state, modes $2$ and $4$ are effectively passive throughout the experiment. So reshaping with $U_x(\tfrac{\theta}{2})$ (equivalent to reshaping with $e^{i\theta(\tilde{n}_1 + \tilde{n}_2)}$ by \eq{number_Ux_equiv}) is not actually necessary. Partial decoupling using $U_x^{(1,3)}(\tfrac{\theta}{2})$ (equivalent to using $e^{i\theta \tilde{n}_1}$) would suffice. The latter is a two-mode unitary acting on modes $1$ and $3$. On the other hand, $U_x (\frac{\theta}{2}) = U_x^{(1,3)}(\frac{\theta}{2})U_x^{(2,4)}(\frac{\theta}{2})$, is a product of two-mode unitaries, collectively acting on all four modes, which might be less preferable to implement in certain experimental setups. In the ancilla-free protocol, this choice is not available; we will see shortly that reshaping must be performed with the full $U_x (\frac{\theta}{2})$ operator in that approach.

To learn the coefficient of $\tilde{n}_2$ using the ancilla, the protocol is analogous, with operators $V(-\frac{\pi}{4})$ and $W(-\frac{\pi}{4})$ now acting on system mode $2$ and the ancilla:
\begin{equation}
\begin{aligned}
    V(\theta) &= e^{\theta (\create{a}{2}\create{b}{} - \ann{b}{}\ann{a}{2})}\\
    W(\theta) &= e^{i \theta (\create{a}{2}\create{b}{} + \ann{b}{}\ann{a}{2})}.
\end{aligned}
\end{equation}
The initial state is $U_y^{(2,4)}(\tfrac{\pi}{4})V(-\frac{\pi}{4})\ket{\Omega} = \frac{1}{\sqrt{2}}(\ket{\Omega} + \create{\tilde{a}}{2}\create{b}{}\ket{\Omega})$. Running RPE yields an estimate of Re($h_{24}$), assuming $\omega_2$ and $\omega_4$ have been previously estimated.

\subsubsection*{b. Learning hopping coefficients: ancilla-free protocol}
Alternatively, to learn Re($h_{13}$) and Re($h_{24}$) without the ancilla, we utilize not just the previously estimated chemical potentials $\omega_1,\ldots,\omega_4$, but also the on-site interaction coefficients $\xi_{12}$ and $\xi_{34}$. By using two different initial states, we can estimate linear combinations of these eight parameters, from which estimates of Re($h_{13}$) and Re($h_{24}$) can be extracted.

For the first round of experiments, we use slightly modified versions of the operators $V(-\frac{\pi}{4})$ and $W(-\frac{\pi}{4})$ defined in \eq{local_unitaries_def}, now acting on system modes $1$ and $2$:
\begin{equation}
\begin{aligned}
    V(\theta) &= e^{\theta (\create{a}{1}\create{a}{2} - \ann{a}{2}\ann{a}{1})}\\
    W(\theta) &= e^{i \theta (\create{a}{1}\create{a}{2} + \ann{a}{2}\ann{a}{1})}.
\end{aligned}
\end{equation}

Defining $U_y (\theta) = U_y^{(1,3)}(\theta)\, U_y^{(2,4)}(\theta)$, the initial state used is $U_y(\tfrac{\pi}{4})V(-\frac{\pi}{4})\ket{\Omega} = \frac{1}{\sqrt{2}}(\ket{\Omega} + \create{\tilde{a}}{1}\create{\tilde{a}}{2}\ket{\Omega})$. In this ancilla-free setting, reshaping must be performed with $U_x(\tfrac{\theta}{2})$ to ensure that all hopping terms are eliminated. Defining $p = \frac{\omega_1 + \omega_3}{2}$, $q = \frac{\omega_2 + \omega_4}{2}$, and $r = \frac{\xi_{12}+\xi_{34}}{4}$, the Type-0 success probability (that of obtaining the vacuum state upon measurement) is given by
\begin{equation}
\begin{aligned}
   p_0 &= |\bra{\Omega}V^{\dag}(-\tfrac{\pi}{4})U_y^{\dag}(\tfrac{\pi}{4})\biggl(\prod_{j=1}^r U_x\biggl(\frac{\theta_j}{2}\biggr) e^{-iHt/r} U_x\biggl(-\frac{\theta_j}{2}\biggr)\biggr) U_y(\tfrac{\pi}{4})V(-\tfrac{\pi}{4})\ket{\Omega}|^2\\  
   &= \frac{1}{2}\biggl(1+\cos\bigl(p+q+r + \text{Re}(h_{13}) + \text{Re}(h_{24})\bigr)\biggr) + \delta_0
\end{aligned}
\end{equation}
The Type-+ success probability is given by
\begin{equation}
\begin{aligned}
   p_+ &= |\bra{\Omega}W^{\dag}(-\tfrac{\pi}{4})U_y^{\dag}(\tfrac{\pi}{4})\biggl(\prod_{j=1}^r U_x\biggl(\frac{\theta_j}{2}\biggr) e^{-iHt/r} U_x\biggl(-\frac{\theta_j}{2}\biggr)\biggr) U_y(\tfrac{\pi}{4})V(-\tfrac{\pi}{4})\ket{\Omega}|^2\\  
   &= \frac{1}{2}\biggl(1+\sin\bigl(p+q+r + \text{Re}(h_{13}) + \text{Re}(h_{24})\bigr)\biggr) + \delta_+.
\end{aligned}
\end{equation}
Using RPE, these experiments provide an estimate of $(p+q+r + \text{Re}(h_{13}) + \text{Re}(h_{24}))$.

For the next round of experiments, we use different initial states and final measurement bases. To define them, we use the operators $V'(-\frac{\pi}{4})$ and $W'(-\frac{\pi}{4})$ defined in \eq{local_unitaries_def_2}. The initial state for RPE is $U_y(\tfrac{\pi}{4})V'(-\tfrac{\pi}{4})(\create{a}{1}\ket{\Omega}) = \frac{1}{\sqrt{2}}(\create{\tilde{a}}{1}\ket{\Omega} + \create{\tilde{a}}{2}\ket{\Omega})$. This is analogous to the state used in the ancilla-free approach to learning $\omega_2$ in \sect{single_site_oscillator} (the state in the first line of \eq{local_unitaries_action_2}), but in the $U_y$-basis. Reshaping with $U_x(\tfrac{\theta}{2})$ and applying $V'^{\dag}(-\tfrac{\pi}{4})U_y^{\dag}(\tfrac{\pi}{4})$ before measurement, the Type-0 success probability (that of obtaining $\create{a}{1}\ket{\Omega}$ upon measurement) is given by:
\begin{equation}
\begin{aligned}
   p_0 &= |(\bra{\Omega} \ann{a}{1}) V'^{\dag}(-\tfrac{\pi}{4})U_y^{\dag}(\tfrac{\pi}{4})\biggl(\prod_{j=1}^r U_x\biggl(\frac{\theta_j}{2}\biggr) e^{-iHt/r} U_x\biggl(-\frac{\theta_j}{2}\biggr)\biggr) U_y(\tfrac{\pi}{4})V'(-\tfrac{\pi}{4})(\create{a}{1}\ket{\Omega})|^2\\  
   &= \frac{1}{2}\biggl(1+\cos\bigl(\text{Re}(h_{24}) - \text{Re}(h_{13}) + q-p\bigr)\biggr) + \delta_0
\end{aligned}
\end{equation}
Similarly, applying $W'^{\dag}(-\tfrac{\pi}{4})U_y^{\dag}(\tfrac{\pi}{4})$ before measurement, the Type-+ success probability is given by
\begin{equation}
\begin{aligned}
   p_+ &= |(\bra{\Omega} \ann{a}{1}) W'^{\dag}(-\tfrac{\pi}{4})U_y^{\dag}(\tfrac{\pi}{4})\biggl(\prod_{j=1}^r U_x\biggl(\frac{\theta_j}{2}\biggr) e^{-iHt/r} U_x\biggl(-\frac{\theta_j}{2}\biggr)\biggr) U_y(\tfrac{\pi}{4})V'(-\tfrac{\pi}{4})(\create{a}{1}\ket{\Omega})|^2\\  
   &= \frac{1}{2}\biggl(1+\sin\bigl(\text{Re}(h_{24}) - \text{Re}(h_{13}) + q-p\bigr)\biggr) + \delta_+.
\end{aligned}
\end{equation}
Using RPE, these experiments provide an estimate of $(\text{Re}(h_{24}) - \text{Re}(h_{13}) + q-p)$.

Now, using the previous estimates of $\omega_1,\omega_2, \omega_3, \omega_4, \xi_{12}$ and $\xi_{34}$, we have estimates of $p, q$ and $r$. By taking the appropriate linear combinations of our newly obtained estimates for $(p+q+r + \text{Re}(h_{13}) + \text{Re}(h_{24}))$ and $(\text{Re}(h_{24}) - \text{Re}(h_{13}) + q-p)$, we can estimate $\text{Re}(h_{13})$ and $\text{Re}(h_{24})$.\footnote{One subtlety to note here is that learning these linear combinations modulo $2\pi$ using RPE would only allow one to infer $2\text{Re}(h_{13})$ and $2\text{Re}(h_{24})$ mod $2\pi$, hence $\text{Re}(h_{13})$ and $\text{Re}(h_{24})$ mod $\pi$. This ambiguity is easily overcome by modifying the evolution times in each experiment of the RPE protocol by a factor of $\frac{1}{2}$, resulting in estimates of $\text{Re}(h_{13})$ and $\text{Re}(h_{24})$ mod $2\pi$ as desired.}

This concludes the discussion of the two approaches to learning the real components, $\text{Re}(h_{13})$ and $\text{Re}(h_{24})$, of the hopping coefficients. Finally, to learn the imaginary components Im($h_{13}$) and Im($h_{24}$), we can use the same procedures as for the real components, but with the roles of $U_x$ and $U_y$ interchanged. 
The transformed fermionic operators $\tilde{a}_1, \tilde{a}_2, \tilde{a}_3, \tilde{a}_4$ in this case are given by
\begin{equation}\label{eq:mode_transfs_Ux}
\begin{aligned}
    \begin{pmatrix}
         \tilde{a}_1 \\
         \tilde{a}_3
     \end{pmatrix}
    &= 
    \begin{pmatrix}
         U_x^{(1,3)}(\tfrac{\pi}{4})\, a_1\, U_x^{\dag (1,3)}(\tfrac{\pi}{4})\\
         U_x^{(1,3)}(\tfrac{\pi}{4}) \,a_3\, U_x^{\dag (1,3)}(\tfrac{\pi}{4})
     \end{pmatrix}
     = \frac{1}{\sqrt{2}}\begin{pmatrix}
         1 & -i\\
         -i & 1
     \end{pmatrix}
     \begin{pmatrix}
         a_1\\
         a_3
     \end{pmatrix}\\
     \begin{pmatrix}
         \tilde{a}_2 \\
         \tilde{a}_4
     \end{pmatrix}
    &= 
    \begin{pmatrix}
         U_x^{(2,4)}(\tfrac{\pi}{4})\, a_2\, U_x^{\dag (2,4)}(\tfrac{\pi}{4})\\
         U_x^{(2,4)}\tfrac{\pi}{4}) \,a_4\, U_x^{\dag (2,4)}(\tfrac{\pi}{4})
     \end{pmatrix}
     = \frac{1}{\sqrt{2}}\begin{pmatrix}
         1 & -i\\
         -i & 1
     \end{pmatrix}
     \begin{pmatrix}
         a_2\\
         a_4
     \end{pmatrix}.
\end{aligned}
\end{equation}
The basis-transformed Hamiltonian has a similar form as \eq{two_site_H_Uy}, with the coefficient of $\tilde{n}_1$ being $\frac{\omega_1 + \omega_3}{2} - \text{Im}(h_{13})$ and the coefficient of $\tilde{n}_2$ being $\frac{\omega_2 + \omega_4}{2} - \text{Im}(h_{24})$, hence it suffices to learn these two coefficients. The unitaries used to reshape the Hamiltonian can be expressed in terms of $U_y (\theta) = U_y^{(1,3)}(\theta)\, U_y^{(2,4)}(\theta)$, observing that $e^{i\theta(\tilde{n}_1 + \tilde{n}_2)} = e^{i\frac{\theta}{2}(n_1+n_2+n_3+n_4)} U_y(\tfrac{\theta}{2})$. The rest of the procedure, whether ancilla-based or ancilla-free, is analogous to the procedure for the real components.

\subsection{Learning a many-body Hamiltonian}\label{sec:many_body_H}
Having considered the single-site and two-site cases, we finally come to the $N$-site many-body Hamiltonians of interest:
\begin{equation}\label{eq:manybody_H}
    H = \sum_{<i,j>} \sum_{\sigma \in \{\uparrow, \downarrow\}} h_{ij \sigma} \create{a}{i \sigma}\ann{a}{j \sigma} +  \sum_{\sigma \in \{\uparrow, \downarrow\}} \omega_{i\sigma} n_{i\sigma} + \sum_i \xi_i n_{i \uparrow} n_{i \downarrow}.
\end{equation}
As in \sect{model_and_results}, we are now explicitly indexing the \textit{spatial} sites by $i,j$, and distinguishing the two spin modes at each site by $\uparrow$ and $\downarrow$. Pairs of spatial sites coupled by a hopping interaction are denoted by $\langle i,j \rangle$. We can associate each spatial site to a vertex of a graph $G$, with edges between sites coupled by a hopping interaction. We assume that the degree of each vertex is $\mathcal{O}(1)$.

\begin{figure}[h]
\centering
\includegraphics[width = 0.65\linewidth]{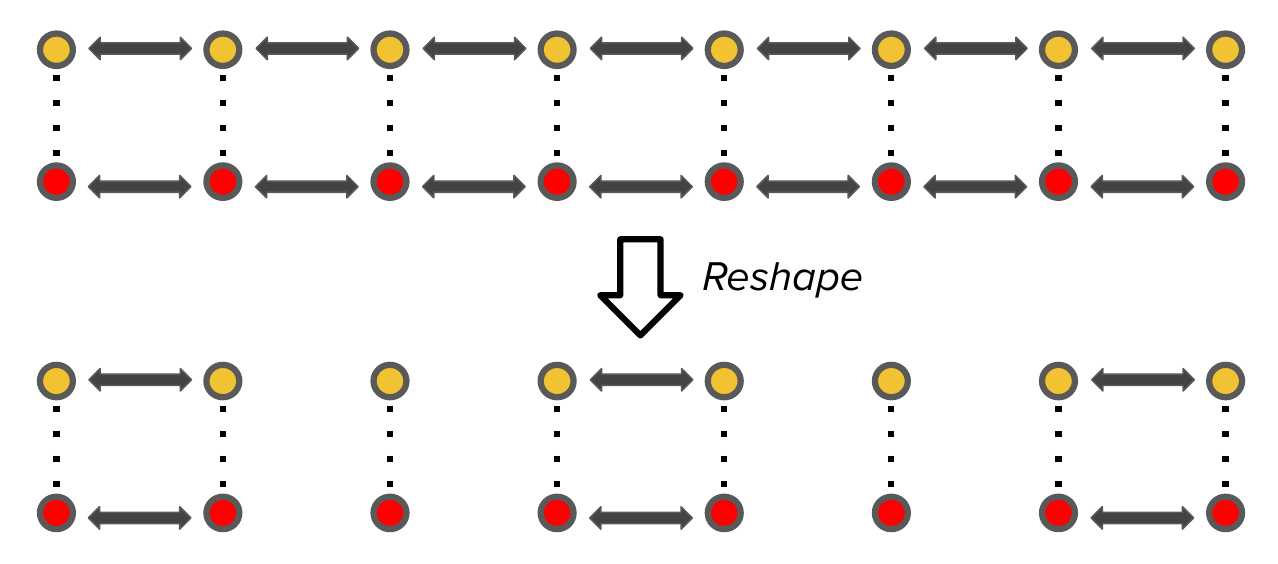}
\caption{By effectively eliminating a subset of the hopping interactions, reshaping decouples the system into independent two-site (four-mode) clusters that can be learned in parallel using the techniques of \sect{two_coupled_oscillators}. This figure depicts reshaping for the special case of a one-dimensional lattice.}
\label{fig:reshaping}
\end{figure}

Hamiltonians of this form can be learned using the same divide-and-conquer strategy used for many-body qubit Hamiltonians \citep{Huang_qubit_2023} and bosonic Hamiltonians \citep{Li_boson_2023}. The key idea is that the decoupling technique used to eliminate hopping terms in the two-site case (\sect{two_coupled_oscillators}) can be used to decouple the Hamiltonian in \eq{manybody_H} into multiple non-interacting clusters, whose parameters can be learned in parallel. The learning problem is then effectively reduced to the two-site case. A simple example of this in the case of a one-dimensional lattice is depicted in \fig{reshaping}.

To perform the decoupling, we start with the graph $G = (V, E)$ of fermionic sites. $V$ is the set of vertices (representing spatial fermionic sites) and $E$ is the set of edges. Using the terminology of \citep{Li_boson_2023}, consider the \textit{link graph}, $L = (E, E_L)$. The set of vertices of the link graph $L$ is the set $E$ of edges of the original graph $G$, so each vertex of $L$ represents a pair $\langle i,j \rangle$ of spatial sites coupled by a hopping interaction. There is an edge between two vertices of $L$ if and only if the corresponding edges in the original graph $G$ intersect at a vertex. In other words, the interactions that they represent involve a common spatial site. The link graph $L$ can be viewed as the analogue of the \textit{dual interaction graph} of qubit Hamiltonians \citep{Huang_qubit_2023, Haah_2023}. 

The link graph $L$ is used to partition the Hamiltonian into clusters that can be decoupled. Specifically, consider a coloring of $L$ such that each vertex is a different color from all its neighbors and next-nearest neighbors. The number of colors needed is $\chi$. As noted in \citep{Li_boson_2023}, $\chi \leq 4(\text{deg}(G)-1)^2 + 1$, and such a coloring can be found by a greedy algorithm. The coloring partitions the vertices of $L$ into disjoint sets $E_c$, each labelled by a different color $c \in \{1,\dots,\chi\}$.

For each color $c$, the goal is to reshape the Hamiltonian such that the interactions $E\backslash E_c$ are eliminated. That is, the only surviving interactions are the ones represented by $E_c$. As in \sect{two_coupled_oscillators}, this can be achieved by conjugating the Hamiltonian by suitably distributed random unitaries. Let $V_c$ be the set of spatial sites (vertices of $G$) involved in the interactions represented by $E_c$. The unitaries used for reshaping have the form 
\begin{equation}\label{eq:reshaping_op_manybody}
    \prod_{j \in V \backslash V_c} e^{-i\theta_j(n_{j \uparrow} + n_{j \downarrow})}
\end{equation}
with each $\theta_j \sim \mathcal{U}([0,2\pi])$.
By similar calculations to those used in \sect{two_coupled_oscillators}, one can check that the Hamiltonian obtained by conjugating with the unitaries above eliminates, in expectation, all but those interactions contained in $E_c$:
\begin{equation}\label{eq:reshaped_manybody_H}
    H = \sum_{<i,j> \in E_c} \sum_{\sigma \in \{\uparrow, \downarrow\}} h_{ij \sigma} \create{a}{i \sigma}\ann{a}{j \sigma} +  \sum_{\sigma \in \{\uparrow, \downarrow\}} \omega_{i\sigma} n_{i\sigma} + \sum_i \xi_i n_{i \uparrow} n_{i \downarrow}
\end{equation}

Furthermore, by construction (due to the coloring rules for the link graph), each pair of interacting vertices $\langle i,j \rangle$ in the reshaped Hamiltonian has no overlap with any other pair of interacting vertices. Thus, the reshaped Hamiltonian is a sum of decoupled two-site Hamiltonians, which can be learned in parallel using the approach described in \sect{two_coupled_oscillators}. In fact, as explained in \sect{single_site_oscillator} and \sect{two_coupled_oscillators}, some of the unitaries in \eq{reshaping_op_manybody} can be replaced by either $e^{-i\theta_j n_{j \uparrow}}$ or $e^{-i\theta_j n_{j \downarrow}}$ in the ancilla-based protocols, depending on which coefficient is to be learned. We can ignore the terms $\omega_{i\sigma} n_{i\sigma}$ and $\xi_i n_{i \uparrow} n_{i \downarrow}$ for $i \in V\backslash V_c$ in the reshaped Hamiltonian, since they have no effect due to our choice of initial states.

Repeating this process for each one of the $\chi$-many colors $c$ results in learning all the parameters of the original Hamiltonian. Therefore, the time complexity overhead compared to the two-site case is a multiplicative factor of $\chi$. Since the graph $G$ has bounded degree, $\chi$ and hence the time complexity of the procedure remains $\order{1/\varepsilon}$, independent of the number of fermionic sites $N$. To implement ancilla-based protocols for any given any color, two ancillae are needed for each of the $\order{N}$ two-mode subsystems being learned in parallel, leading to an overhead of $\order{N}$ ancillae. The ancillae can be re-used across multiple experiments, so the ancilla count does not depend on the number of experimental rounds or number of colors.

To summarize the conceptual structure of the entire protocol:
\begin{enumerate}
    \item Use a coloring of the link graph defined above to determine which two-site clusters can be learned in parallel. That is, determine which hopping interactions are to be eliminated by reshaping in order to obtain effectively independent two-site clusters. Each set of independent clusters corresponds to one of the colors used in the graph coloring. (The coloring is used to identify the appropriate unitary distribution for reshaping. The reshaping itself occurs during time evolution in each experiment, by inserting unitaries sampled from this distribution after sufficiently small time-steps of Hamiltonian evolution, as explained in \sect{two_coupled_oscillators}).
    \item For each color, learn the Hamiltonian parameters for all the independent two-site clusters in parallel, by performing the following sequence of steps on each cluster (comprising spatial sites $i$ and $j$):
        \begin{enumerate}
            \item Learn $\omega_{i\uparrow}$ and $\omega_{j\uparrow}$ in parallel (two ancillae required), as described in \sect{number_conserving_coeffs}.
            \item Learn $\omega_{i\downarrow}$ and $\omega_{j\downarrow}$ in parallel (ancillae not required given odd-parity initial states), as described in \sect{number_conserving_coeffs}.
            \item Learn $\xi_{i}$ and $\xi_{j}$ in parallel (no ancillae), as described in \sect{number_conserving_coeffs}.
            \item Learn Re($h_{ij\uparrow}$) and Re($h_{ij\downarrow}$) either sequentially (one ancilla required) or in parallel (no ancillae), as described in \sect{learn_hopping_coeffs}.
            \item Learn Im($h_{ij\uparrow}$) and Im($h_{ij\downarrow}$) either sequentially (one ancilla required) or in parallel (no ancillae), as described in \sect{learn_hopping_coeffs}.
        \end{enumerate}
\end{enumerate}

\section{Discussion}\label{sec:discussion}
In this paper, we have addressed the problem of learning a class of fermionic Hubbard Hamiltonians of physical interest, with complex hopping amplitudes, nonzero chemical potentials, and on-site interactions. We have shown that the parameters of such Hamiltonians can be learned at the Heisenberg limit, where the total evolution time across all experiments is $\mathcal{O}(1/\varepsilon)$ and the number of experiments is $\mathcal{O}(\text{polylog}(1/\varepsilon))$, as long as the graph representing the fermionic interactions has bounded degree. Each experiment involves preparing fermionic Gaussian states, alternating time evolution with fermionic linear optics (FLO) unitaries according to the paradigm of discrete quantum control, performing local occupation number measurements on the fermionic modes, and classically post-processing the results according to the Robust Phase Estimation algorithm \citep{RPE_Kimmel_2015, RPE_Russo_2021, RPE_Belliardo_2020}. Some of the experiments utilize $\order{N}$ fermionic ancilla modes, where $N$ is the system size, to satisfy the constraints imposed by the parity superselection rule for fermions. The protocol is robust to a constant amount of state preparation and measurement error.

We saw that for certain steps of the protocol, the use of ancillae is optional, but provides a tradeoff in terms of resources. To learn some of the chemical potential terms, if odd-parity initial states are available, the ancillae enable discrete quantum control to be effective with single-mode FLO unitaries instead of two-mode FLO unitaries. In more physical terms, this means that ancillae allow trading two-mode fermionic phase-shifters for single-mode phase-shifters. While learning the hopping coefficients, ancillae enable the use of two-mode unitaries rather than a product of two-mode unitaries on non-overlapping modes -- in other words, trading two beamsplitters for a single beamsplitter. The use of ancillae, for those steps of the protocol where they are optional, may therefore depend on which operations are easier to implement on a given platform. It's easy to imagine other minor variations on the approach presented here that would also achieve the Heisenberg limit. For application to a specific platform, there is flexibility to tune the protocol to use the operations most easily implemented there.

A natural next step would be to find a Heisenberg-limited algorithm for our class of fermionic Hubbard Hamiltonians that eliminates the ancilla overhead altogether, for instance by learning appropriate linear combinations of the Hamiltonian parameters from which the individual parameters can be inferred. Furthermore, although both the bosonic Hamiltonians considered in \citep{Li_boson_2023} and the fermionic Hamiltonians considered here are physically well-motivated, they are still special cases of the larger set of physically interesting Hamiltonians. For instance, one could consider Hamiltonians that include mix of bosonic, fermionic and qubit sites, such as the Bose-Fermi Hubbard model \citep{Bukov_2014} and the Jaynes-Cummings model \citep{Jaynes_1963}. An interesting next step would be to investigate whether Heisenberg-limited learning is achievable for such Hamiltonians. More generally, however, it would be desirable to have a simple criterion, applying to a wide range of Hamiltonians, that serves as a sufficient condition for Heisenberg-limited learning to be possible without depending on system size. This is the case for qubit Hamiltonians \citep{Huang_qubit_2023, Dutkiewicz_2023}, where the relevant criterion is that the Hamiltonian is low-intersection. This criterion encompasses a larger set of qubit Hamiltonians compared to the more restricted classes of bosonic and fermionic Hamiltonians for which Heisenberg-limited learning algorithms have been established. A useful next step would therefore be to identify an assumption analogous to the low-intersection assumption, to extend these results to a wider set of bosonic and fermionic Hamiltonians.

\begin{acknowledgments}
We thank Vedika Khemani, Sri Raghu and Nicole Ticea for discussions on the Hubbard model, Di Fang, Jeongwan Haah and Yu Tong for discussions about qubit Hamiltonian learning algorithms, and Lexing Ying for introducing us to the bosonic case. This work was supported by the Herb and Jane Dwight Stanford Graduate Fellowship, ARO (award W911NF2120214), DOE (Q-NEXT), CIFAR and the Simons Foundation.
\end{acknowledgments}

\appendix

\section{Analysis of reshaping error}\label{appendix:app_a}
In this section, we consider in more detail the error due to the approximate time evolution operator obtained via the reshaping procedure, introduced in \sect{number_conserving_coeffs}:
\begin{equation}\label{eq:approx_op}
    \prod_{j=1}^r U_j^{\dag} e^{-iHt/r} U_j
\end{equation}
We wish to demonstrate that the following bound holds for the expectation values of any operator $O_S$ with support on an $\order{1}$ subset $S$ of fermionic modes and $\norm{O_S} \leq 1$:
\begin{equation}\label{eq:op_error_bound_2}
    \left| \text{Tr}(\rho(t)_{\text{approx}} O_S) - \text{Tr}(\rho(t)_{\text{exact}} O_S)\right| \leq \order{\frac{t^2}{r}}
\end{equation}
where
\begin{equation}\rho(t)_{\text{approx}} = \mathbb{E}_{U_j \sim \mathcal{D}}\biggl(\displaystyle{\prod_{1\leq j \leq r}^{\leftarrow}} U_j^{\dag} e^{-iHt/r} U_j\biggr) \rho \biggl(\prod_{1\leq j \leq r}^{\rightarrow} U_j^{\dag} e^{iHt/r} U_j\biggr),
\end{equation}
and $\rho(t)_{\text{exact}} = e^{-iH_{\text{eff}}t} \rho e^{iH_{\text{eff}}t}$. Here $H_{\text{eff}} = \mathbb{E}_{U \sim \mathcal{D}} U^{\dag} H U = H'_S + H'_{S^c}$, that is, the effective Hamiltonian decouples the original Hamiltonian into a sum of a Hamiltonian supported on $S$ and a Hamiltonian supported on the rest of the system $S^c$. $O_S$ acts nontrivially only on the modes in $S$, and acts implicitly as the identity operator on the other modes. The big-$\mathcal{O}$ notation in \eq{op_error_bound_2} means specifically that there is a constant independent of the system size $N$.

The outcome probabilities $p_0$ and $p_+$ of the various RPE experiments outlined in \sect{two_coupled_oscillators} can manifestly be expressed as expectation values of projection operators, which have norm $1$ and act on at most four modes. Hence these projectors satisfy the requirements for the operators $O_S$ in \eq{op_error_bound_2}, and \eq{op_error_bound_2} justifies the use of the approximate time evolution operator in \eq{approx_op} for the RPE experiments. Throughout this section we set the largest absolute value among the Hamiltonian coefficients, $\lambda_{max} = 1$. (For any other finite value, all the complexities in this section would simply pick up a constant multiplicative factor of $\lambda^2_{max}$).

The bound in \eq{op_error_bound_2} is a corollary of the following bound originally derived for the special case of qubit Hamiltonians by Huang et al. \citep{Huang_qubit_2023}: 
\begin{equation}\label{eq:qubit_error_bound}
    \norm{\mathbb{E}_{U_j \sim \mathcal{D}}\biggl(\prod_{1\leq j \leq r}^{\rightarrow} U_j^{\dag} e^{iHt/r} U_j\biggr) O_S \biggl(\prod_{1\leq j \leq r}^{\leftarrow} U_j^{\dag} e^{-iHt/r} U_j\biggr) - e^{iH_{\text{eff}}t} O_S e^{-iH_{\text{eff}}t}} \leq \order{\frac{t^2}{r}}
\end{equation}
Since $H_{\text{eff}} = H'_S + H'_{S^c}$ and $O_S$ acts nontrivially only on $S$, the   $H'_{S^c}$ term does not contribute to the Heisenberg-picture evolution of $O_S$. Hence the second term on the left hand side of $\eq{qubit_error_bound}$ represents the ideal subsystem projector for RPE that acts nontrivially only on $S$, while the first term is the approximate projector that appears in the actual protocol. 

As observed by \citep{Ni_fermion}, the proof of \eq{qubit_error_bound} in \citep{Huang_qubit_2023} applies to more general unitaries than the tensor products of Pauli operators considered in \citep{Huang_qubit_2023}, including those relevant for the protocol of \citep{Ni_fermion} and our protocol in this paper. Hence \eq{qubit_error_bound} can be proved analogously to the proof technique used in \citep{Huang_qubit_2023}.

\section{Root-mean-square error of linear combination of estimators}\label{appendix:app_b}
In each of the ancilla-free protocols described in \sect{results}, the RPE procedure does not directly yield an estimate of the desired parameter. Rather, it estimates a linear combination of the desired parameter and other parameters that have previously been estimated. We then estimate the desired parameter by appropriately adding or subtracting the previously obtained estimators of the other parameters in the linear combination. This process is straightforward when the estimators being linearly combined are obtained from different experiments and therefore independent, as in the single-site Hamiltonian of \sect{single_site_oscillator}. In this case, if the RMS error of each individual summand is $\leq \order{\varepsilon}$ (for the Heisenberg-scaling $\varepsilon$ of RPE), the RMS error of the linear combination is also $\leq \order{\varepsilon}$, since Var($X+Y$) = Var($X$) + Var($Y$) for independent random variables \citep{probability_textbook}.

However, in the ancilla-free subroutines of \sect{two_coupled_oscillators}, to learn the coefficients of a two-site Hamiltonian, the estimators being combined are not always independent. In particular, pairs of parameters are learned in parallel via the decoupling technique ($\omega_1$ with $\omega_3$, $\omega_2$ with $\omega_4$, and $\xi_{12}$ with $\xi_{34}$). If decoupling were perfect, then the estimators obtained in parallel would be independent. In practice, since decoupling is approximate (as discussed in Appendix A), the estimators of parameters learned in parallel are not truly independent. Nevertheless, by the Cauchy-Schwarz inequality, the covariance of any two random variables $X$ and $Y$, Cov($X,Y$), is bounded by Cov($X,Y$) $\leq$ Var($X$) + Var($Y$) \citep{probability_textbook}. Additionally, Var($X+Y$) = Var($X$) + Var($Y$) + Cov($X,Y$). Since each of the estimators being combined has RMS error $\leq \order{\varepsilon}$, these facts imply that their linear combinations also have RMS error $\leq \order{\varepsilon}$ as desired.

\bibliography{references}

\end{document}